\documentclass[twocolumn,secnumarabic,amssymb,nobibnotes,aps,superscriptaddress,pra]{revtex4-1}

\usepackage[dvips]{graphicx}
\usepackage{bm}
\usepackage{latexsym} 
\usepackage{amsmath}
\usepackage{amssymb}
\usepackage{color}

\newcommand{\ui}{{\rm i}}
\newcommand{\veps}{{\varepsilon}}

\newcommand{\bmr}{{\bm r}}
\newcommand{\bmp}{{\bm p}} 

\newcommand{\bmq}{{\bm q}}
\newcommand{\bra}{\langle}
\newcommand{\ket}{\rangle}
\newcommand{\kB}{k_{\rm B}}

\makeatletter
\renewcommand*{\p@subsection}{}
\renewcommand*{\p@subsubsection}{}
\makeatother

\begin{document}

\title{
  Spin diffusion equation in superconductors in the vicinity of $T_{\rm c}$ 
}

\author{Takuya Taira} 
\affiliation{Department of Physics, Okayama University, Okayama 700-8530, Japan}
\author{Masanori Ichioka}
\affiliation{Research Institute for Interdisciplinary Science, Okayama University, Okayama 700-8530, Japan}
\affiliation{Department of Physics, Okayama University, Okayama 700-8530, Japan}
\author{So Takei} 
\affiliation{
Department of Physics, Queens College of the City University of New York, Queens, New York 11367, USA}

\author{Hiroto Adachi}
\affiliation{Research Institute for Interdisciplinary Science, Okayama University, Okayama 700-8530, Japan}
\affiliation{Department of Physics, Okayama University, Okayama 700-8530, Japan}
\date{\today}

\begin{abstract} 
  We microscopically derive the spin diffusion equation in an $s$-wave superconductor in the vicinity of the superconducting transition temperature $T_{\rm c}$. Applying the general relation between the relaxation function and the response function to the present spin diffusion problem, we examine how the spin relaxation time and the spin diffusion coefficient are renormalized in the superconducting state. The analysis reveals that, below $T_{\rm c}$, both the spin relaxation time and the spin diffusion coefficient are increased, resulting in an enhancement of the spin diffusion length. The present result may provide an explanation for the recent observation of an enhanced spin pumping signal below $T_{\rm c}$ in a Py/Nb/Pt system that is free from the coherence peak effect. 
\end{abstract}

\pacs{}

\keywords{} 

\maketitle

\section{Introduction \label{Sec:I}}

The interplay of the spin current and superconductivity has been attracting much attention, forming a new research field of superconducting spintronics~\cite{Linder15,Eschrig15}. Historically, the interplay of spin and superconductivity has been studied for a long time, and it is well known that the static perturbation having the {\it odd} time-reversal symmetry, such as spin, affects the thermodynamic properties of an $s$-wave superconductor~\cite{Anderson59,Abrikosov-Gorkov,Maki-review,Balatsky-review}. By contrast, the spin current is a flow of spin and thus has the {\it even} time-reversal symmetry~\cite{Rashba06,Nagaosa08}. Therefore, the interplay of spin current and superconductivity~\cite{Bell08,Yang10,Hubler12,Quay13,Wakamura14,Ohnishi14,Yao18,Umeda18} has an intriguing issue from the viewpoint of extending Anderson's theorem~\cite{Anderson59} to dynamics, i.e., it requires to clarify the effects of {\it dynamic} and {\it even time-reversal symmetry} perturbation on the superconductivity. 

When investigating such spin current physics, the most basic theoretical apparatus is the spin diffusion equation. Since after Bloembergen~\cite{Bloembergen49} discussed the diffusion of nuclear spins in a crystalline solid, this equation has been applied to a number of different physical situations~\cite{Torrey56,Walker71,Aronov76,Silsbee79,vanSon87,Johnson88,Valet93,Hershfield97,Schmidt00,Rashba00,Yu02,Takahashi03}. The advent of the spin transfer torque concept~\cite{Slonczewski96,Berger96} has further reminded us of the importance of the spin diffusion equation. Indeed the real space profile of the spin current is described by this equation, which can now be experimentally measured by the so-called lateral spin valve technique~\cite{Jedema01}. Therefore, in order to investigate the interplay of spin current and superconductivity, it is of great importance to know the spin diffusion equation in the superconducting state. Despite its importance, however, only a little is known about the spin diffusion equation in the superconducting state~\cite{Yafet83,Yamashita02}. 

In this work, we microscopically derive the spin diffusion equation in the superconducting state on the basis of the weak-coupling BCS model with $s$-wave pairing and with impurity spin-orbit scattering. Nevertheless, the derivation is not an easy task as one can infer from the fact that the superconductivity involves a nontrivial many-body ground state, and that we need to deal with the nonequilibrium problem in discussing the spin current. To overcome these difficulties, we limit ourselves to a temperature region near $T_{\rm c}$, and derive the spin diffusion equation by perturbation with respect to the superconducting gap $\Delta$. It is important to note that, compared with the static linear response, the present dynamic linear response problem is much more involved, requiring a precise evaluation of the diffusion pole. Besides, the calculation in the $T \ll T_{\rm c}$ region can be numerically problematic because of the growing gap edge singularity. In order to obtain analytically reliable and transparent result, we focus on a region in the vicinity of $T_{\rm c}$, and the in-depth description deep in the superconducting state is left to future works.

Our microscopic derivation of the spin diffusion equation is enabled by the use of the general relation between the relaxation function and the response function, which in the present context is translated into the relation between the spin diffusion equation and the dynamic spin susceptibility. This relation is well known in the nonequilibrium statistical mechanics~\cite{Kubo-text,Chaikin-text}. Employing the formulation of \cite{Inoue17}, where the dynamic spin susceptibility in the superconducting state was calculated, we derive the spin diffusion equation microscopically. Then, we show that the spin relaxation time as well as the spin diffusion coefficient are increased below $T_{\rm c}$, leading to an increment of the spin diffusion length in the superconducting state.

In order to compare the theory with experiments, we next apply the present result to a quite recent experiment measuring the spin transport through a superconducting Nb~\cite{Jeon18}. In \cite{Jeon18}, the spin pumping in a Py/Nb/Pt system was studied, and it was found that the spin transport across the Nb layer from the Py to Pt layers is enhanced below $T_{\rm c}$. Then it is argued that the enhancement of the spin pumping in Nb below $T_{\rm c}$ is due to a possible formation of triplet Cooper pairs. We analyze the corresponding physical situation using the result derived in the present paper. Then, we provide an alternative explanation for the enhancement in the spin transport below and close to $T_c$ without relying on the spin transport mediated by triplet Cooper pairs. 

This paper is organized as follows. In the next section, we briefly review the derivation of the important relation between the spin diffusion equation and the dynamic spin susceptibility that constitutes the basis of the present work. In Sec.~\ref{Sec:III}, we introduce our model and the Green's function, the latter of which is expanded in powers of the superconducting gap $\Delta$. In Sec.~\ref{Sec:IV}, using the relation between the spin diffusion equation and the spin susceptibility that is obtained in Sec.~\ref{Sec:II}, we derive the spin diffusion equation both in the normal state and in the superconducting state near $T_{\rm c}$. In Sec.~\ref{Sec:V}, relying on the result derived in the present paper, we calculate the spin pumping signal in a situation of Ref.~\cite{Jeon18} in the vicinity of $T_{\rm c}$. Finally, in Sec.~\ref{Sec:VI} we discuss and summarize our results. We use units $\hbar=\kB=1$ throughout this paper.

\section{Relation between spin diffusion equation and spin susceptibility \label{Sec:II}} 

\begin{figure}[t] 
  \begin{center}
    \scalebox{0.35}[0.35]{\includegraphics{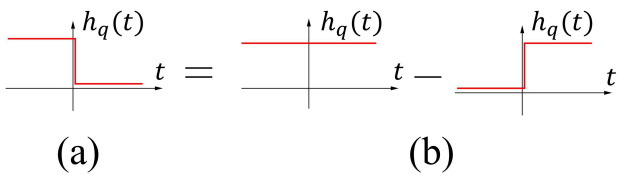}}
  \end{center}
  \caption{Schematic drawing of the external magnetic field $h_\bmq (t)$ applied to the system. (a) Perturbation describing spin diffusion. (b) Decomposition of the same perturbation into two parts, describing static and dynamic responses. 
  }
  \label{fig:relax-resp}
\end{figure}

In this section, we briefly review the derivation of the well-known relation between the relaxation function and the response function~\cite{Kubo-text,Chaikin-text}, which in the present context is translated into the relation between the spin diffusion equation and the dynamic spin susceptibility. In Sec.~\ref{Sec:IV}, this relation will be used to microscopically derive the spin diffusion equation.

We begin with the following spin diffusion equation for the spin density $\bra s(\bmr,t) \ket$: 
\begin{equation}
  \left( \frac{\partial}{\partial t} - D_s \nabla^2 + \frac{1}{\tau_s} \right) \bra s(\bmr,t) \ket =0,
  \label{eq:spin-diffusion01}
\end{equation}
where $D_s$ is the spin diffusion coefficient and $\tau_s$ is the spin relaxation time. Following \cite{Chaikin-text}, we consider a physical situation [Fig.~\ref{fig:relax-resp}(a)] where a spatially non-uniform static external magnetic field $h_\bmq {\bm \hat{\bm z}}$ is applied to the system at time $t<0$ with which the spin density is in equilibrium. Then, the external magnetic field is turned off at $t=0$, and the spin density diffuses for $t>0$ to achieve the spatially uniform equilibrium state. The external perturbation describing this situation can be written as
\begin{equation}
  h_\bmq(t)= \Theta(-t) e^{0_+ t} h_\bmq,
  \label{eq:hq_t01}
\end{equation}
where $\Theta(t)$ and $0_+$ are respectively the step function and positive infinitesimal constant. 

Consider first the time evolution of $\bra s(\bmr,t) \ket$ for $t>0$. Using the momentum representation $\bra s(\bmr,t) \ket= \int d^3 q \bra s_\bmq(t) \ket e^{\ui \bmq \cdot \bmr}$ as well as introducing the Laplace-Fourier transformation $\bra s_\bmq(z) \ket = \int_0^\infty dt \bra s_\bmq(t) \ket e^{\ui z t}$, we obtain the solution to the spin diffusion equation (\ref{eq:spin-diffusion01}):
\begin{equation}
  \bra s_\bmq (z) \ket = R_\bmq (z) \bra s_\bmq(t=0) \ket, 
\end{equation}
where $\bra s_\bmq(t=0) \ket$ is the spin density at time $t=0$, $z= \omega+ \ui 0_+$ is the complex frequency in the upper half-plane to ensure causality, 
and 
\begin{equation}
  R_\bmq (z) = \frac{1}{-\ui z + D_s q^2 + \tau_s^{-1}}
  \label{eq:relaxfunc01}
\end{equation}
is the relaxation function of the spin diffusion equation. Applying the result of static linear response to the initial spin density, i.e., $\bra s_\bmq(t=0) \ket= \chi_\bmq(0) h_\bmq $, we obtain
\begin{equation}
  \bra s_\bmq (z) \ket = R_\bmq (z) \chi_\bmq(0) h_\bmq,
  \label{eq:relaxation01}
\end{equation}
where $\chi_\bmq(0)$ is the static spin susceptibility.

Consider next the dynamic linear response to the external magnetic field represented by Eq.~(\ref{eq:hq_t01}). The basic idea is that the time dependence of the external magnetic field can be decomposed into the static field and the step-function field [Fig.~\ref{fig:relax-resp}(b)]. Since the static magnetic field defines the static spin susceptibility $\chi_\bmq(0)$ and the step-function field defines the dynamic (retarded) spin susceptibility $\chi^R_\bmq (z)$, we can relate $\bra s_\bmq (z) \ket$ in Eq.~(\ref{eq:relaxation01}) with $\chi_\bmq(0)$ and $\chi^R_\bmq (z)$. Indeed, for the external magnetic field of Eq.~(\ref{eq:hq_t01}), we have the relation 
\begin{equation}
  \bra s_\bmq (t) \ket= \int_{-\infty}^0 dt' \; \chi^R_\bmq(t-t') e^{0_+ t'} h_\bmq. 
  \label{eq:dynamic01}
\end{equation}
Then, going into the frequency space and using the spectral representation
\begin{equation}
  \chi^R_\bmq (z) =
  \int_{-\infty}^\infty \frac{d \omega'}{\pi} \frac{{\rm Im} \chi^R_\bmq(\omega')}{\omega'- z - \ui 0_+},
  \label{eq:spectral01}
\end{equation}
Eq.~(\ref{eq:dynamic01}) can be rewritten as
\begin{equation}
  \bra s_\bmq (z) \ket = 
  \int_{-\infty}^\infty \frac{d \omega'}{\ui \pi}
  \frac{{\rm Im} \chi^R_\bmq(\omega') h_\bmq}{(\omega'- z)(\omega'- \ui 0_+)}. 
\end{equation}
After using of the partial fraction decomposition
\begin{equation}
  \frac{1}{(\omega'- z)(\omega'-\ui 0_+)} =
  \frac{1}{z} \left( \frac{1}{\omega'- z}- \frac{1}{\omega'-\ui 0_+} \right),
\end{equation}  
we arrive at 
\begin{equation}
  \bra s_\bmq (z) \ket =
  \frac{1}{\ui z} \Big( \chi^R_\bmq(z)- \chi_\bmq(0) \Big) h_\bmq,
    \label{eq:relaxation02}
\end{equation}
where we again used the spectral representation (\ref{eq:spectral01}).

Now we compare Eqs.~(\ref{eq:relaxation01}) and (\ref{eq:relaxation02}), and obtain the key equation: 
\begin{equation}
  R_\bmq(z) =
  \frac{1}{\ui z} \left( \frac{\chi^R_\bmq(z)}{\chi_\bmq(0)} -1 \right), 
  \label{eq:relaxation03}
\end{equation}
which relates the spin diffusion equation with the dynamic spin susceptibility. Moreover, we use Eq.~(\ref{eq:relaxfunc01}) by setting $z= \omega+ \ui 0_+$ and arrive at the following identity: 
\begin{equation}
  -\ui \omega + D_s q^2+ \frac{1}{\tau_s} = 
  \frac{\ui \omega}{\left( \frac{\chi^R_\bmq(\omega)}{\chi_\bmq(0)} -1 \right)},
  \label{eq:spin-diffusion02}
\end{equation}
where the small $\omega$, small $q^2$ limit should be taken. Eq.~(\ref{eq:spin-diffusion02}) means that spin diffusion equation can be derived microscopically by calculating the dynamic spin susceptibility $\chi^R_\bmq(\omega)$.

\section{Model \& Green's function near $T_{\rm c}$ \label{Sec:III}} 

In this section, we first introduce our model Hamiltonian and define the corresponding Green's function which is expanded in powers of the superconducting gap $\Delta$. Then, following \cite{Inoue17}, we develop a formalism to calculate the dynamic spin susceptibility up to $\Delta^2$, by making use of the Green's function technique. As expected, the present result is in line with that of Ref.~\cite{Inoue17} in the limit of small $\Delta$. Indeed, as we will see in Appendix~\ref{Sec:App02}, our calculation can correctly reproduce the appearance of the coherence peak predicted for the spin pumping into superconductors~\cite{Inoue17}. 

  A key in our approach is that the vertex corrections are fully taken into account in calculating the dynamic spin susceptibility. It is well known that for a microscopic calculation of transport quantities, we need to keep a certain relation between the single-particle and two-particle Green's functions in order to satisfy the conservation laws~\cite{Baym-Kadanoff61}. In the present formulation, this corresponds to the relation between the selfenergy function and the vertex function, which is usually referred to as Ward identity~\cite{Nozieres-text}. This relation should be kept in the diffusion problem as well~\cite{Maleev75}, where it is known that only after taking account of the vertex corrections can we derive the correct diffusion equation that satisfies the charge conservation. 

As a starting model, we consider the following BCS Hamiltonian for an impure $s$-wave superconductor~\cite{Inoue17}: 
\begin{eqnarray}
  {\cal H} &=&  \sum_{\bmp} c^\dag_{\bmp} \xi_\bmp  c_{\bmp} 
  - \sum_{\bmp} \Big( \Delta^* c_{\bmp \downarrow} c_{-\bmp \uparrow} 
  + \Delta c^\dag_{-\bmp \downarrow} c^\dag_{\bmp  \uparrow} \Big)+ {\cal H}_{\rm imp}, \nonumber \\
  \label{eq:BCS01}
\end{eqnarray}
where $c^\dag_{\bmp}= (c^\dag_{\bmp \uparrow}, c^\dag_{\bmp \downarrow})$ is the electron creation operator for spin projection $\uparrow$ and $\downarrow$, $\xi_\bmp$ is the kinetic energy measured from the chemical potential, and $\Delta$ is the superconducting gap. The last term in Eq.~(\ref{eq:BCS01}), 
\begin{eqnarray}
  {\cal H}_{\rm imp} &=&
  \sum_{\bmp,\bmp'} c^\dag_{\bmp} \hat{U}_{\bmp,\bmp'} c_{\bmp'}, 
\end{eqnarray} 
describes the scattering by impurities. Here, 
\begin{equation}
  \hat{U}_{\bmp,\bmp'}
  = u_{0}({\bmp-\bmp'})+ \ui u_{\rm so} ({\bmp-\bmp'})\hat{\bm \sigma} \cdot (\bmp \times \bmp')
\end{equation}
is the impurity potential, where $\hat{\bm \sigma}$ is the Pauli matrices, $u_{0}({\bmp-\bmp'})$ and $u_{\rm so}({\bmp-\bmp'})$ are the amplitudes of the momentum and spin-orbit scatterings, respectively.

As in \cite{Inoue17,Abrikosov62} we use the representation for the impurity-averaged Green's function, 
\begin{equation}
  \check{G}_\bmp (\ui \veps_n) =   
  \begin{pmatrix}
    {\cal G}_\bmp (\ui \veps_n) , & {\cal F}_\bmp (\ui \veps_n) \ui \hat{\sigma}^y \\
    {\cal F}^\dag_\bmp (\ui \veps_n)\ui \hat{\sigma}^y , & {\cal G}^\dag_\bmp (\ui \veps_n) 
  \end{pmatrix},
  \label{eq:singleG01}
\end{equation}
where $\veps_n= 2 \pi T (n+1/2)$ is a fermionic Matsubara frequency with $n$ being an integer, and we deal with the impurity spin-orbit scattering following Ref.~\cite{Abrikosov62}. We denote a matrix in the particle-hole (Nambu) space by ``check'' accent. Then, the Matsubara susceptibility is obtained from 
\begin{equation}
  \chi_\bmq(\ui \omega_\nu) = - \frac{T}{2} \sum_{\veps_n} \int_\bmp \;
      {\rm Tr} \Big[ \hat{\sigma}^z \hat{\Upsilon}_{\bmp, \bmp_-} (\ui \veps_n, \ui \veps_{n_{-}} ) \Big],
      \label{eq:chiQW01}
\end{equation}
where $\omega_\nu= 2 \pi T \nu$ is a bosonic Matsubara frequency with $\nu$ being an integer, and we have introduced the shorthand notation $\int_\bmp= \int \frac{d^3 p}{(2 \pi)^3}$. Here, $\hat{\Upsilon}_{\bmp, \bmp_-} (\ui \veps_n, \ui \veps_{n_-} )$ is the $(1,1)$ component of the matrix function defined by 
\begin{equation}
  \check{\Upsilon}_{\bmp,\bmp_-}(\ui \veps_n,\ui \veps_{n_-}) = \check{G}_{\bmp}(\ui \veps_n)
  \check{\Lambda}_{\bmq} (\ui \veps_n,\ui \veps_{n_-}) \check{G}_{\bmp_-}(\ui \veps_{n_-}), 
  \label{eq:delG_QW01}
\end{equation}
where $\bmp_-= \bmp - \bmq$ and $\veps_{n_-}= \veps_{n} -\omega_\nu$. In the above equation, the vertex function $\check{\Lambda}_{\bmp,\bmp_-} (\ui \veps_n, \ui \veps_{n_-})$ arises from the effects of impurity ladder, and it satisfies the following equation: 
\begin{eqnarray}
  \check{\Lambda}_{\bmq} (\ui \veps_n,\ui \veps_{n_-}) &=&
  \hat{\sigma}^z + n_{\rm imp} \int_{\bmp'} \check{V}_{\bmp,\bmp'}
  \check{G}_{\bmp'} (\ui \veps_n)  \nonumber \\
  && \times \check{\Lambda}_\bmq (\ui \veps_n,\ui \veps_{n_-}) 
  \check{G}_{\bmp'_-}(\ui \veps_{n_-}) \check{V}_{\bmp',\bmp}, 
  \label{eq:Lambda02}
\end{eqnarray}
where the effect of impurities is described by 
\begin{equation}
  \check{V}_{\bmp,\bmp'} =
  \begin{pmatrix}
    \hat{U}_{\bmp,\bmp'}, & 0 \\
    0, & \hat{U}^t_{\bmp',\bmp}
  \end{pmatrix}, 
\end{equation}
with $\hat{U}^t$ being the transpose of a matrix $\hat{U}$ in the spin space.

In order to solve Eq.~(\ref{eq:Lambda02}), we introduce the representation~\cite{Gorkov64}:
\begin{equation}
  \check{\Lambda}_{\bmq} (\ui \veps_n, \ui \veps_{n_-}) =
  \begin{pmatrix} 
    \Lambda_1 \hat{\sigma}^z,
    & \Lambda_2\hat{\sigma}^x \\
    -\Lambda^\dag_2 \hat{\sigma}^x,
    & \Lambda^\dag_1 \hat{\sigma}^z 
  \end{pmatrix}. 
  \label{eq:Lambda_matrix01}
\end{equation}
In the normal state where $\Delta=0$, since $\check{G}_{\bmp} (\ui \veps_n)$ and $\check{G}_{\bmp_-} (\ui \veps_{n_-})$ commute as they are diagonal in the Nambu space, the relation $\Lambda_1= \Lambda^\dag_1$ and $\Lambda_2= \Lambda^\dag_2=0$ holds. In the superconducting state, if the static limit ($\omega_\nu = 0$) is concerned, since $\check{G}_{\bmp} (\ui \veps_n)$ and $\check{G}_{\bmp_-} (\ui \veps_{n_-})$ coincide and thus commute, we have the important symmetry 
\begin{equation}
  \Lambda_1= \Lambda^\dag_1, \qquad \Lambda_2= \Lambda^\dag_2. 
  \label{eq:Lambda12} 
\end{equation}
Because we are interested in the small $\Delta$ region, we keep this symmetry and focus on a small correction to the spin diffusion equation in the normal state. Note that in the present approach the impurity spin-orbit scattering preserves the spin symmetry of the single-particle Green's function [Eq.~(\ref{eq:singleG01})], but it modifies the spin symmetry of the two-particle Green's function [Eqs.~(\ref{eq:delG_QW01}) and (\ref{eq:Lambda_matrix01})] through the vertex corrections. 

Using Eqs.~(\ref{eq:Lambda_matrix01}) and (\ref{eq:Lambda12}), we can transforms Eq.~(\ref{eq:Lambda02}) into a set of linear equations for $\Lambda_1$ and $\Lambda_2$: 
\begin{eqnarray}
  \begin{pmatrix}
    \Lambda_1 \\    
    \Lambda_2
  \end{pmatrix}
  &=&
  \begin{pmatrix}
    1\\
    0 
  \end{pmatrix}
  +
  \begin{pmatrix}
    {\cal A}, & -{\cal B} \\
    {\cal C}, & {\cal D} 
  \end{pmatrix}
  \begin{pmatrix}
    \Lambda_1 \\    
    \Lambda_2 
  \end{pmatrix},   
  \label{eq:Lambda12v02}
\end{eqnarray}
where the coefficients ${\cal A}$, ${\cal B}$, ${\cal C}$, ${\cal D}$ are calculated from Eq.~(\ref{eq:Lambda02}) as 
\begin{eqnarray}
  {\cal A} &=& g\int_\bmp \left\{ {\cal G}_\bmp(\ui \veps_n) {\cal G}_{\bmp_-} (\ui \veps_{n_-})
  + {\cal F}_\bmp(\ui \veps_n){\cal F}^{\dag}_{\bmp_-} (\ui \veps_- ) \right\}, \label{eq:calA01}\\
  {\cal B} &=& g \int_\bmp \left\{ {\cal G}_\bmp(\ui \veps_n) {\cal F}^{\dag}_{\bmp_-} (\ui \veps_{n_-})
  + {\cal F}_\bmp (\ui \veps_n){\cal G}_{\bmp_-}(\ui \veps_{n_-}) \right\}, \label{eq:calB01}\\
  {\cal C} &=& g \int_\bmp \left\{ {\cal G}_\bmp(\ui \veps_n) {\cal F}_{\bmp_-} (\ui \veps_{n_-})
  - {\cal F}_\bmp(\ui \veps_n){\cal G}^{\dag}_{\bmp_-}(\ui \veps_-) \right\}, \label{eq:calC01}\\
  {\cal D} &=& g \int_\bmp \left\{ {\cal G}_\bmp(\ui \veps_n) {\cal G}^{\dag}_{\bmp_-} (\ui \veps_{n_-})
  - {\cal F}_\bmp(\ui \veps_n){\cal F}_{\bmp_-}(\ui \veps_-) \right\}. \label{eq:calD01}
\end{eqnarray}
Here, $g= {\Gamma_{(-)}}/{\pi N(0)}$, and $N(0)$ is the density of states per one spin projection. The scattering rate $\Gamma_{(-)}$ arising from the vertex corrections is given by 
\begin{equation}
  \Gamma_{(-)} = \frac{1}{2} \left( \frac{1}{\tau_0}- \frac{1}{3 \tau_{\rm so}} \right), 
  \label{eq:Gamma_-01} 
\end{equation}
where $\tau_0$ and $\tau_{\rm so}$ are respectively the momentum relaxation time and spin-orbit relaxation time, which are assumed to satisfy $1/\tau_0 \gg 1/\tau_{\rm so}$ and given by 
${1}/{\tau_0} = 2 \pi N(0) n_{\rm imp} |u_0(0)|^2$ and
${1}/{\tau_{\rm so}} = 2 \pi N(0) n_{\rm imp} |u_{\rm so}(0)|^2 \bra ({\bmp} \times {\bmp}')^2 \ket_{\rm FS}$. Here, $n_{\rm imp}$ is the number density of impurities, and $\bra \cdots \ket_{\rm FS}$ means the average over the Fermi surface. Substituting the solution of Eq.~(\ref{eq:Lambda12v02}) into Eq.~(\ref{eq:chiQW01}), the Matsubara spin susceptibility is calculated to be 
\begin{eqnarray}
  \chi_\bmq (\ui \omega_\nu) &=&
  -\frac{\pi N(0)}{\Gamma_{(-)}} T \sum_{\veps_n}
  \frac{{\cal A}(1-{\cal D})-{\cal B}{\cal C}}{(1-{\cal A})(1-{\cal D})+{\cal B}{\cal C}}. 
  \label{eq:chiQW03}
\end{eqnarray}

Since we are interested in the region immediately below $T_{\rm c}$, we expand the normal Green's functions in powers of $\Delta$~\cite{Gorkov60}: 
\begin{eqnarray}
  {\cal G}_\bmp(\ui \veps_n) &=&
  {\cal G}^{(0)}_\bmp(\ui \veps_n)+ \delta {\cal G}_\bmp(\ui \veps_n)
  + O \left( \Delta^4 \right), \\
  {\cal G}^\dag_\bmp(\ui \veps_n) &=&
  {\cal G}^{\dag (0)}_\bmp(\ui \veps_n)+ \delta {\cal G}^{\dag}_\bmp(\ui \veps_n)
  + O \left( \Delta^4 \right),
\end{eqnarray}
where
\begin{equation}
  {\cal G}^{(0)}_\bmp(\ui \veps_n)= \frac{1}{\ui \widetilde{\veps}_n- \xi_\bmp}, \qquad
  {\cal G}^{\dag (0)}_\bmp(\ui \veps_n)= -\frac{1}{\ui \widetilde{\veps}_n+ \xi_\bmp}, 
\end{equation}
are the Green's functions per one spin projection in the normal state, 
$\widetilde{\veps}_n= \veps_n \eta $ with
\begin{equation}
  \eta= 1+ \frac{\Gamma_{(+)}}{|\veps_n|},
\end{equation}
and the scattering rate $\Gamma_{(+)}$ is given by 
\begin{equation}
  \Gamma_{(+)} = \frac{1}{2} \left( \frac{1}{\tau_0}+ \frac{1}{\tau_{\rm so}} \right). 
\end{equation}

Now we perform a perturbation expansion with respect to $\Delta$. Following the standard procedure (e.g., \cite{Werthamer-review}), we find that $\delta {\cal G}_\bmp(\ui \veps_n)$ and $\delta {\cal G}^\dag_\bmp(\ui \veps_n)$ are given by 
\begin{eqnarray}
  \delta {\cal G}_\bmp(\ui \veps_n)
  &=&
  \left( {\cal G}^{(0)}_\bmp(\ui \veps_n) \right)^2
  \Big\{ \frac{\ui \Gamma_{(+)} {\rm sgn}(\veps_n)}{2 |\veps_n|^2} 
  - \eta^2 {\cal G}^{\dag (0)}_\bmp(\ui \veps_n) \Big\} \Delta^2 , \nonumber \\
  \\
  \delta {\cal G}^{\dag}_\bmp(\ui \veps_n) 
  &=&
  \left( {\cal G}^{\dag (0)}_\bmp(\ui \veps_n) \right)^2
  \Big\{ \eta^2 {\cal G}^{(0)}_\bmp(\ui \veps_n) 
 - \frac{\ui \Gamma_{(+)} {\rm sgn}(\veps_n)}{2 |\veps_n|^2} \Big\} \Delta^2. \nonumber \\  
\end{eqnarray}
In a similar way the anomalous Green's function, ${\cal F}_\bmp(\ui \veps_n)$, which is equal to ${\cal F}^\dag_\bmp(\ui \veps_n)$ in the Meissner state, is expanded as 
\begin{equation}
  {\cal F}_\bmp(\ui \veps_n) = \delta F_\bmp(\ui \veps_n)+ O \left( \Delta^3 \right), 
\end{equation}
where $\delta F_\bmp(\ui \veps_n)$ is given by

\begin{equation}
  \delta F_\bmp(\ui \veps_n)
  = {\cal G}^{\dag (0)}_\bmp(\ui \veps_n) {\Delta}\eta {\cal G}^{(0)}_\bmp(\ui \veps_n).
\end{equation}

\section{Derivation of the Spin diffusion equation \label{Sec:IV}}

In this section, we derive the spin diffusion equation microscopically. First, we focus on the normal state and demonstrate that our procedure correctly reproduces the well-known expression for the spin diffusion equation in the normal state. Next, we extend the derivation to the superconducting state near $T_{\rm c}$, and show that the spin diffusion length is increased below $T_{\rm c}$. 

\subsection{Normal state}

\begin{figure}[t] 
  \begin{center}
    \scalebox{0.5}[0.5]{\includegraphics{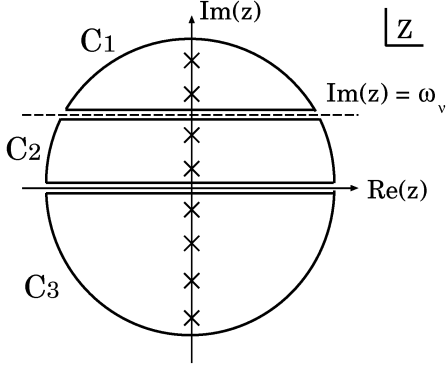}}
  \end{center}
\caption{ 
  Three regions of the summand of the spin susceptibility in the complex $z$ plane, where the Matsubara frequency is located on $z= \ui \veps_n$. 
}
\label{fig1_taira}
\end{figure}

Let us first calculate the Matsubara susceptibility $\chi_\bmq(\ui \omega_\nu)$ in the normal state, by using Eq.~(\ref{eq:chiQW03}). Noting that ${\cal B}= {\cal C}= {\cal D}=0$ in the normal state, the Matsubara susceptibility is calculated to be 
\begin{equation}
  \chi_\bmq(\ui \omega_\nu) = - \frac{\pi N(0)}{\Gamma_{(-)}}
  T \sum_{\veps_n} \frac{{\cal A}^{(0)}}{1-{\cal A}^{(0)}},
  \label{eq:chi_n01}
\end{equation}
where 
\begin{equation}
  {\cal A}^{(0)} = g 
  \int_\bmp {\cal G}^{(0)}_\bmp(\ui \veps_n) {\cal G}^{(0)}_{\bmp-\bmq}(\ui \veps_n- \ui \omega_\nu) 
  \label{eq:A0}
\end{equation}
is the normal state value of ${\cal A}$. The summation over Matsubara frequency in Eq.~(\ref{eq:chi_n01}) requires a care, because the analytic behavior of ${\cal A}^{(0)}$ changes discontinuously depending on the sign of $\veps_n (\veps_n- \omega_\nu)$. Therefore, it is convenient to divide the susceptibility into two parts~\cite{AGD}: 
\begin{equation}
  \chi_\bmq (\ui \omega_\nu) = \chi'_\bmq (\ui \omega_\nu)+ \chi''_\bmq (\ui \omega_\nu),
  \label{eq:chi1chi2matsu}
\end{equation}
where $\chi'_\bmq (\ui \omega_\nu)$ comes from the ``non-dissipative'' region $\veps_n (\veps_n- \omega_\nu)>0$, i.e., 
\begin{equation}
  \chi'_\bmq(\ui \omega_\nu) = - \frac{\pi N(0)}{\Gamma_{(-)}}
  \left( T \sum_{\veps_n<0} + T \sum_{\omega_\nu < \veps_n} \right) \frac{{\cal A}^{(0)}}{1-{\cal A}^{(0)}}, 
\end{equation}
whereas $\chi''_\bmq (\ui \omega_\nu)$ comes from the ``dissipative'' region $\veps_n (\veps_n- \omega_\nu)<0$, i.e., 
\begin{equation}
  \chi''_\bmq(\ui \omega_\nu) = - \frac{\pi N(0)}{\Gamma_{(-)}}
  T \sum_{0< \veps_n < \omega_\nu }\frac{{\cal A}^{(0)}}{1-{\cal A}^{(0)}}.
  \label{eq:chi2matsu01}
\end{equation}
On the complex $z$ plane where the Matsubara frequency is located at $z= \ui \veps_n$ (Fig.~\ref{fig1_taira}), $\chi'_\bmq (\ui \omega_\nu)$ is given by the sum over regions $C_1$ and $C_3$, whereas $\chi''_\bmq (\ui \omega_\nu)$ is given by the sum over region $C_2$.

We next evaluate $\chi'_\bmq (\ui \omega_\nu)$ and $\chi''_\bmq (\ui \omega_\nu)$ in the normal state. Regarding $\chi'_\bmq (\ui \omega_\nu)$, it is well known that this quantity is dominated by the static susceptibility 
\begin{equation}
  \chi'_\bmq (\ui \omega_\nu) \approx N(0), 
\end{equation}
where the result is insensitive to the effects of impurities~\cite{Fulde68}. Regarding $\chi''_\bmq (\ui \omega_\nu)$, since ${\cal A}^{(0)}$ in the ``dissipative'' region is given by 
${\cal A}^{(0)} = \frac{\Gamma_{(-)}}{\Gamma_{(+)}} \left( 1- \frac{\omega_\nu+ D q^2}{2 \Gamma_{(+)}} \right)$ 
with $D= v_F^2 \tau_{\rm tr}/3$ being the diffusion coefficient, we obtain 
\begin{eqnarray}
  \chi''_\bmq(\ui \omega_\nu) 
  &=& -\frac{N(0) \omega_\nu }
        {2\Gamma_{(+)}} \frac{{\cal A}^{(0)}}{1- {\cal A}^{(0)}}, 
        \label{eq:chi2matsu02}
\end{eqnarray}
where in moving from Eq.~(\ref{eq:chi2matsu01}) to Eq.~(\ref{eq:chi2matsu02}), the summation over $\veps_n$ is replaced by multiplication by a factor $\omega_\nu/2 \pi T$, i.e., the number of Matsubara frequencies in the region $0<\veps_n<\omega_\nu$, as the summand is independent of $\veps_n$~\cite{Fukuyama85}.

As the final step, we perform the analytic continuation $\ui \omega_\nu \to \omega+ \ui 0_+$. Then, ${\cal A}^{(0)}$ is calculated to be 
\begin{equation}
  {\cal A}^{(0)} = \frac{\Gamma_{(-)}}{\Gamma_{(+)}} \left( 1- \frac{-\ui \omega + D q^2}{2 \Gamma_{(+)}} \right),
  \label{eq:A0ret}
\end{equation}
from which we obtain the retarded spin susceptibility, 
\begin{equation}
  \chi^{R}_\bmq(\omega)  
  = N(0)
  \frac{Dq^2+ (4/3\tau_{\rm so})}{-\ui \omega + Dq^2+ (4/3\tau_{\rm so})}, 
  \label{eq:chi_n_R01}
\end{equation}
due to Eq.~(\ref{eq:chi2matsu02}). Substituting the above equation into Eq.~(\ref{eq:spin-diffusion02}), we obtain the spin diffusion coefficient
\begin{equation}
  D^{({\rm n})}_s = \frac{v_F^2 \tau_{\rm tr}}{3}, 
\end{equation}
and the spin relaxation time 
\begin{equation}
  \tau^{({\rm n})}_s = \frac{3}{4} \tau_{\rm so}, 
\end{equation}
where the superscript ${\rm (n)}$ means the value in the normal state.

\subsection{Superconducting state near $T_{\rm c}$}

We now calculate the spin diffusion equation in the vicinity of the superconducting transition, and show that the spin diffusion length $\lambda_s = \sqrt{D_s \tau_s}$ is increased below $T_{\rm c}$. Since we are interested in an analytic result valid near $T_{\rm c}$, we expand the dynamic spin susceptibility up to $\Delta^2$.

First, we consider the ``non-dissipative'' component, $\chi'_\bmq (\ui \omega_\nu)$, which is dominated by the uniform susceptibility derived in \cite{Abrikosov62}. The result can be expanded up to $\Delta^2$ (see Appendix~\ref{Sec:App01}), 
\begin{eqnarray}
  \chi'_\bmq (\ui \omega_\nu) &=&
  N(0) \left\{ 1- 2 \pi T \sum_{\veps_n > 0} \frac{\Delta^2}{\veps_n^3+ \frac{2}{3 \tau_{\rm so}}} \right\}
  + O(\Delta^4) \nonumber \\
  &=& N(0) \left\{ 1- \left( \frac{7\zeta(3)}{4}- \frac{5 \zeta(4)}{4 \veps_0 \tau_{\rm so}} \right)
  \frac{\Delta^2}{\veps_0^2}\right\} + O(\Delta^4), \nonumber \\
  \label{eq:Rechi01}
\end{eqnarray}
where $\veps_0$ is the lowest fermionic Matsubara frequency, 
\begin{equation}
  \veps_0= \pi T.
  \label{eq:veps_0}
\end{equation}

Next, we consider the ``dissipative'' region $\veps_n (\veps_n- \omega_\nu)<0$. To calculate $\chi''_\bmq (\ui \omega_\nu)$ up to $\Delta^2$, we expand the coefficients ${\cal A}$ in powers of $\Delta$: 
\begin{eqnarray}
  {\cal A} &=& {\cal A}^{(0)}+ \delta {\cal A}+ O \left( \Delta^4 \right),
  \label{eq:dA01}
\end{eqnarray}
where ${\cal A}^{(0)}$ is defined by Eq.~(\ref{eq:A0}). In the above equation, $\delta {\cal A}$ is given by 
\begin{eqnarray}
  \delta {\cal A}  &=&
  g \int_\bmp \Big\{ \delta {\cal G}(p) {\cal G}^{(0)} (p_-) 
  + {\cal G}^{(0)}(p) \delta {\cal G} (p_-)\nonumber \\
  && + \delta {\cal F} (p)\delta {\cal F}^{\dag} (p_-) 
  + \delta {\cal F}(p) \delta {\cal F}^{\dag} (p_-)\Big\},  
\end{eqnarray}
where we have introduced the four-vector notation $p= (\ui \veps_n, \bmp)$, namely, $\delta {\cal G}(p) = \delta {\cal G}_\bmp(\ui \veps_n)$, ${\cal G}^{(0)}(p_-) = {\cal G}^{(0)}_{\bmp_-}(\ui \veps_{n_-})$, etc. Recalling that the normal state values of ${\cal B}$ and ${\cal C}$ vanish, we expand ${\cal B}$ and ${\cal C}$ as follows: 
\begin{eqnarray}
  {\cal B} &=& \delta {\cal B} + O \left( \Delta^3 \right),
  \label{eq:dB01}\\
        {\cal C} &=& \delta {\cal C} + O \left( \Delta^3 \right),
        \label{eq:dC01}
\end{eqnarray}
where ${\cal B}$ and ${\cal C}$ are both of the first order with respect to $\Delta$, which are explicitly given by 
\begin{eqnarray}
  \delta {\cal B}  &=& 
  g \int_\bmp \left\{ {\cal G}^{(0)}(p)\delta {\cal F}(p_-) + \delta {\cal F}(p){\cal G}^{(0)}(p_-) \right\}, \\
  \delta {\cal C} &=& 
  g \int_\bmp \left\{ {\cal G}^{(0)}(p)\delta {\cal F}(p_-) - \delta {\cal F}(p){\cal G}^{\dag (0)}(p_-) \right\}, 
\end{eqnarray}
where $g$ is defined below Eq.~(\ref{eq:calD01}). 

Finally, the coefficient ${\cal D}$ in the ``dissipative'' region is expanded as 
\begin{eqnarray}
  {\cal D} &=& \delta {\cal D} + O \left( \Delta^4 \right),
  \label{eq:dD01}
\end{eqnarray}
where, as we will see below, only the fact that $\delta {\cal D} \sim O(\Delta^2)$ is important and the detailed expression of $\delta {\cal D}$ is unnecessary.

With these apparatus, let us calculate $\chi''_\bmq(\ui \omega_\nu)$. Substituting Eqs.~(\ref{eq:dA01}), (\ref{eq:dB01}), (\ref{eq:dC01}), and (\ref{eq:dD01}) into Eq.(\ref{eq:chiQW03}), and expand the numerator and the denominator up to $\Delta^2$, we obtain 
\begin{equation}
  \chi''_\bmq (\ui \omega_\nu) =
  -\frac{\pi N(0)}{\Gamma_{(-)}} T \sum_{0< \veps_n<\omega_\nu}
  {\cal S} (\veps_n) 
  \label{eq:Imchi01}
\end{equation}
where the summand is defined by 
\begin{equation}
  {\cal S} (\veps_n) =
  \frac{{\cal A}^{(0)}+    \left[\delta {\cal A} 
      -\delta {\cal B} \delta {\cal C} \right]}
       {1-{\cal A}^{(0)} - \left[\delta {\cal A} 
           -\delta {\cal B} \delta {\cal C} \right]}.
       \label{eq:S01}
\end{equation} 
To proceed further, we first recall that the summand is independent of $\veps_n$ in the normal state, and that we are considering a small correction to the normal state, of the order of $\sim (\Delta/T)^2 \ll 1$. Under this circumstances, the summand depends very weakly on $\veps_n$. Therefore we regard the summand as nearly independent of $\veps_n$ in the ``dissipative'' region $0<\veps_n<\omega_\nu$, and approximate 
\begin{equation}
  T \sum_{0< \veps_n<\omega_\nu} {\cal S} (\veps_n) \approx
  T \frac{\omega_\nu}{2 \pi T} {\cal S} (\veps_0),
  \label{eq:summand01} 
\end{equation}
where $\veps_0$ is defined by Eq.~(\ref{eq:veps_0}). Note that this approximation is justified in the case with sizable impurity spin-orbit scattering where the unperturbed term of the denominator in Eq.~(\ref{eq:S01}), $1- {\cal A}^{(0)}$, acquires a nonvanishing value $3/4\tau_{\rm so}$ even in the limit $\omega, Dq^2 \to 0 $. Then, we obtain the expression 
\begin{equation}
  \chi''_\bmq (\ui \omega_\nu) = 
  -\frac{N(0) \omega_\nu}{2 \Gamma_{(-)}}
  \frac{{\cal A}^{(0)}+    \left[\delta {\cal A} 
      -\delta {\cal B}  \delta {\cal C} \right]}
       {1-{\cal A}^{(0)} - \left[\delta {\cal A} 
          -\delta {\cal B}  \delta {\cal C}   \right]},
       \label{eq:Imchi02}
\end{equation}
where the quantities $\delta {\cal A}$, $\delta {\cal B}$, and $\delta {\cal C}$ are now evaluated at $\veps_n= \veps_0$. Keep this in mind and using approximation consistent with Eq.~(\ref{eq:Lambda12}), after a bit tedious algebra, these quantities are calculated to be 
\begin{eqnarray}
  \delta {\cal A} &=& -\frac{\Gamma_{(-)}}{\Gamma_{(+)}}
  \frac{\Delta^2}{\veps_0^2} 
  \frac{\gamma+ (2- 3 \gamma) \frac{Dq^2}{2\Gamma_{(+)}}}{(1-\gamma)^2}, \label{eq:dA02}\\
  \delta {\cal B} &=& \delta {\cal C} 
  = -\ui \frac{\Gamma_{(-)}}{\Gamma_{(+)}} \frac{\Delta}{\veps_0} 
  \frac{1- \frac{Dq^2}{2\Gamma_{(+)}}}{1- \gamma }, \label{eq:dBC02}
\end{eqnarray}
where $\gamma = \Gamma_{(+)}/\veps_0$. Note that $\gamma$ should be in the region $\gamma < 1$ to obtain physical results (see Appendix~\ref{Sec:App02}). 

We are now in a position to calculate the retarded susceptibility. In the following, we normalize the susceptibility by the density of states in the normal state as
\begin{equation}
  \chi^R_\bmq(\omega) = N(0) \widetilde{\chi}^R_\bmq (\omega). 
\end{equation}
Starting from Eqs.~(\ref{eq:Rechi01}) and (\ref{eq:Imchi02}), then performing analytic continuation $\ui \omega_\nu \to \omega + \ui 0_+$, we obtain
\begin{eqnarray}
  \widetilde{\chi}^R_\bmq(\omega) &=&
  1+ \delta \widetilde{\chi}_0 + \frac{\ui \omega}{2 \Gamma_{(-)}}
      \frac{{\cal A}^{(0)}+   \delta {\cal E}}
           {1-{\cal A}^{(0)} - \delta {\cal E} },
           \label{eq:chiR01}
\end{eqnarray}
where
\begin{equation}
  \delta \widetilde{\chi}_0 = - \left( \frac{7\zeta(3)}{4}- \frac{5 \zeta(4)}{4 \veps_0 \tau_{\rm so}} \right)
  \frac{\Delta^2}{\veps_0^2}. 
\end{equation}
In Eq.~(\ref{eq:chiR01}), ${\cal A}^{(0)}$ is given by Eq.~(\ref{eq:A0ret}), and $\delta {\cal E}$ is defined by 
\begin{equation}
  \delta {\cal E} = \delta {\cal A}-\delta {\cal B}  \delta {\cal C}. 
\end{equation}
From Eqs.~(\ref{eq:dA02}) and (\ref{eq:dBC02}), $\delta {\cal E}$ is calculated to be 
\begin{eqnarray}
  \delta {\cal E} &=& \delta {\cal E}_0+ \delta {\cal E}_1 q^2, \label{eq:E00}\\
  \delta {\cal E}_0 &=& \frac{\Delta^2}{\veps_0^2} \frac{1}{1- \gamma},  \label{eq:E0}\\
  \delta {\cal E}_1 &=& -\frac{\Delta^2}{\veps_0^2} \frac{D}{2 \Gamma_{(+)}}
  \frac{4- 3 \gamma}{(1- \gamma)^2}, \label{eq:E1}
\end{eqnarray}
where we have used $\Gamma_{(+)}/\Gamma_{(-)} \approx 1$ in this small correction term. Using the above equation as well as Eq.~(\ref{eq:A0ret}), we can represent the retarded susceptibility in the following form: 
\begin{eqnarray}
  \widetilde{\chi}^R_\bmq(\omega) &=& 
  \frac{(\delta {\cal E}_0 -\delta \widetilde{\chi}_0) \ui \omega + (1+\delta \widetilde{\chi}_0)
    \left(D_s q^2+ {1}/{\tau_s}\right)}
       {-\ui \omega + D_sq^2 + {1}/{\tau_s} },
       \label{eq:chi_SC01}
\end{eqnarray}
where we have introduced $D_s= D_s^{\rm (n)}- 2 \Gamma_{(+)} \delta {\cal E}_1$ and 
${1}/{\tau_s} = {1}/{\tau_s^{\rm (n)}}- 2 \Gamma_{(+)} \delta {\cal E}_0 $. Using Eqs.~(\ref{eq:E0}) and (\ref{eq:E1}), $D_s$ and $\tau_s$ can be expressed as 
\begin{eqnarray}
  D_s &\approx&  D^{\rm (n)}_s \left(1+ \frac{\Delta^2}{\veps_0^2}\frac{4 -3 \gamma}{(1- \gamma)^2} \right), \label{eq:Ds01}\\
  \tau_s &\approx& 
  \tau^{\rm (n)}_s \left( 1 + 
  \frac{\Delta^2}{\veps_0^2}\frac{2 \Gamma_{(+)} \tau^{\rm (n)}_s }{1- \gamma}\right) \label{eq:taus01}. 
\end{eqnarray}

Now we substitute Eq.~(\ref{eq:chi_SC01}) into Eq.~(\ref{eq:spin-diffusion02}) to derive the spin diffusion equation. Then, we obtain the following equation: 
\begin{equation}
  \frac{1+\delta \widetilde{\chi}_0}{1+ \delta {\cal E}_0} \left( -\ui \omega + D_s q^2 + \frac{1}{\tau_s} \right)= 0, 
\end{equation}
from which we find the following spin diffusion equation: 
\begin{equation}
  \left( \frac{\partial}{\partial t} - D_s \nabla^2 + \frac{1}{\tau_s} \right) \bra s(\bmr,t) \ket =0,
  \label{eq:SDE01}
\end{equation}
where $D_s$ and $\tau_s$ are defined by Eqs.~(\ref{eq:Ds01}) and (\ref{eq:taus01}).

The result thus obtained means that, in the superconducting state near $T_{\rm c}$, $D_s $ [Eq.~(\ref{eq:Ds01})] is identified as the spin diffusion coefficient, and $\tau_s$ [Eq.~(\ref{eq:taus01})] is identified as the spin relaxation time. Note that both quantities are increased upon the superconducting transition. Finally, the spin diffusion length is calculated to be 
\begin{eqnarray}
\lambda_s &=& \sqrt{D_s \tau_s} \nonumber \\
&\approx& \lambda^{\rm (n)}_s 
\left( 1+ \frac{\Delta^2}{\veps_0^2} \left[ \frac{\Gamma_{(+)} \tau^{\rm (n)}_s }{1-\gamma}
+ \frac{4- 3 \gamma}{2(1-\gamma)^2} \right]
\right) ,
\label{eq:lambda01}
\end{eqnarray}
where $\lambda^{\rm (n)}_s = \sqrt{D^{\rm (n)}_s \tau^{\rm (n)}_s}$. Therefore, the spin diffusion length is also increased below $T_{\rm c}$. 

\section{Application to experiments \label{Sec:V}} 
In this section, we perform a model calculation to compare the present result with experiments. We apply the theory to a quite recent experiment observing an enhanced spin pumping into the superconducting Nb in a Py/Nb/Pt structure~\cite{Jeon18}. In \cite{Jeon18}, the observed enhancement of the spin pumping signal is attributed to a putative spin supercurrent carried by triplet Cooper pairs. Below, relying on the spin diffusion carried by quasiparticles, we analyze the corresponding physical situation and put forward an alternative interpretation of the experiment near $T_{\rm c}$ without invoking the triplet spin supercurrent. 

\begin{figure}[t] 
  \begin{center}
    \scalebox{0.5}[0.5]{\includegraphics{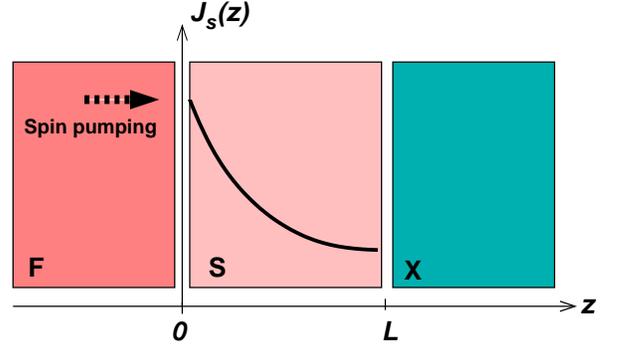}}
  \end{center}
  \caption{Schematic view of the spin current profile in the F/S/X system discussed in Sec.\ref{Sec:V}, where F is a ferromagnet, S is an $s$-wave superconductor, and X is either a vacuum or a perfect spin sink.} 
  \label{fig_FSsink}
\end{figure}

As a simple model calculation, we consider an F/S/X system as shown in Fig.~\ref{fig_FSsink}, where F and S are a ferromagnet and an $s$-wave superconductor, respectively, and X is either a vacuum or a perfect spin sink. Then, we assume that the distribution of the spin accumulation $\delta s$ in $S$ is described by the spin diffusion equation [Eq.~(\ref{eq:SDE01})]. Since the lateral dimensions of the multilayer ($5$~mm $\times$ $5$~mm) are much larger than the layer thickness ($\sim$ several tens of nanometers)~\cite{Jeon18}, we assume an approximate translational invariance along the lateral directions. Therefore, the spin accumulation $\delta s$ and the spin current $J_s$ vary only in the thickness direction, which we take as $z$-axis. Then, the spin accumulation is written as 
\begin{equation}
  \delta s(z) = A \sinh \left( \frac{z}{\lambda_s}\right) + B \cosh \left( \frac{z}{\lambda_s}\right),  
\end{equation}
where $A$, and $B$ are unknown coefficients to be determined by the boundary conditions. The spin current inside S, $J_s(z) = - D_s \nabla_z \delta s (z)$, is given by 
\begin{equation}
  J_s(z) = -\frac{D_s}{\lambda_s}
  \left\{ A \cosh \left( \frac{z}{\lambda_s}\right) + B \sinh \left( \frac{z}{\lambda_s}\right)
  \right\}.
\end{equation}

Now, following \cite{ZhangZhang12}, we consider the boundary conditions. At the F/S interface, we impose 
\begin{equation}
  J^{\rm tot}_s= J_s^{\rm pump}- J_s^{\rm back}, 
  \label{eq:BC01}
\end{equation}
where the first term on the right-hand side is the pumped spin current whereas the second term is the backflow spin current. Here, the pumped current is given by $J_s^{\rm pump}= G_s \delta m(0)$, where $G_s$ corresponds to the so-called spin mixing conductance with the driving force being represented in the form of nonequilibrium magnon accumulation $\delta m(0)$ at the interface~\cite{Kajiwara10}. The backflow spin current is given by $J_s^{\rm back}= G'_s \delta s (0)$, where $G'_s$ is a coefficient proportional to $G_s$. 

Next, we consider the boundary condition at the S/X interface. When X is a vacuum, because the spin current does not penetrate into the vacuum, we have the boundary condition 
\begin{equation}
  J_s(L) = 0 
  \label{eq:BC02}
\end{equation}
at the S/X interface. From the two boundary conditions [Eqs.~(\ref{eq:BC01}) and (\ref{eq:BC02})], we obtain 
\begin{equation}
  J^{\rm tot}_s= \frac{G_s \delta m(0)}{1+ \left(\frac{\lambda_s G'_s}{D_s}\right) 
    \coth \left(\frac{L}{\lambda_s}\right)}.
  \label{eq:Js01}
\end{equation} 
On the other hand, when X is a perfect spin sink, we have the boundary condition 
\begin{equation}
  \delta s(L) = 0,
  \label{eq:BC03}
\end{equation}
and using the two boundary conditions [Eqs.~(\ref{eq:BC01}) and (\ref{eq:BC03})], we have 
\begin{equation}
  J^{\rm tot}_s = \frac{G_s \delta m(0)}{1+ \left(\frac{\lambda_s G'_s}{D_s}\right)
    \tanh \left(\frac{L}{\lambda_s}\right)}. 
  \label{eq:Js02}   
\end{equation}
In the following, we set $G'_s = G_s$.

In order to discuss temperature dependence of $J^{\rm tot}_s$ below $T_{\rm c}$, we need to know temperature dependences of $G_s$, $D_s$, and $\lambda_s$. Limiting ourselves to a narrow temperature region just below $T_{\rm c}$ as well as using the BCS expression $\Delta^2/\veps_0^2= 0.95(T_{\rm c}- T)/T_{\rm c}$ in the mean-field approximation~\cite{AGD}, we parametrize these temperature dependences as 
\begin{eqnarray}
  G_s(T) &=& G_s(T_{\rm c}) \left( 1- a \frac{T_{\rm c}- T}{T_{\rm c}} \right), \\
  D_s(T) &=& D_s(T_{\rm c}) \left( 1+ b \frac{T_{\rm c}- T}{T_{\rm c}} \right), \\
  \lambda_s(T) &=& \lambda_s(T_{\rm c}) \left( 1+ c \frac{T_{\rm c}- T}{T_{\rm c}} \right), 
\end{eqnarray}
where $({T_{\rm c}- T})/{T_{\rm c}}$ measures the distance from $T_{\rm c}$, and three dimensionless parameters $a$, $b$, $c$ give the slopes. Note that in a Py/Pt system, it is reported that $G_s$ decreases below $T_{\rm c}$~\cite{Bell08,Jeon18} presumably because of the too strong exchange interaction at the interface, causing a local destruction of the superconducting gap there~\cite{Tokuyasu88}. Thus, the sign in front of $a$ is set positive. By contrast, $D_s$ and $\lambda_s$ increase below $T_{\rm c}$ according to the present work, such that the sign in front of $b$ and $c$ are selected positive. 

In our model calculation we use $a=1.0$, $b= 2.0$, and $c= 3.0$. This is because $\lambda_s$ is expected to have the strongest temperature dependence due to the largeness of the parameter $\Gamma_{(+)} \tau_s \gg 1$ in Eq.~(\ref{eq:lambda01}). Besides, $G_s$ has a competition between the reduced quasiparticle density of states and the coherence peak effect~\cite{Inoue17,Yao18,Umeda18} and is thus expected to have the weakest temperature dependence. Besides, we use following parameters~\cite{Jeon18}: $L= 30$nm, $\lambda_s(T_{\rm c})= 40$nm, $\lambda_s(T_{\rm c}) G_s(T_{\rm c})/D_s(T_{\rm c})= \lambda_s (T_{\rm c}) g_s/\sigma= 3.2$ calculated from $g_s= e^2 N(0) G_s = 1\times 10^{15}$m$^{-2}\Omega^{-1}$ and $\sigma= e^2 N(0)D_s(T_{\rm c}) =1.25 \times 10^{7} \Omega^{-1}$m$^{-1}$. Note that $J_s^{\rm tot}$ corresponds to the spin pumping signal measured in experiments.

Within the parameter space we have investigated, there are three competing factors in the temperature dependence of $J^{\rm tot}_s$: First, a reduction in $G_s$ results in a slight decrease of $J^{\rm tot}_s$ below $T_{\rm c}$. Second, an increase in $D_s$ results in an enhancement of $J^{\rm tot}_s$ below $T_{\rm c}$. These two tendencies apply irrespective of the choice of boundary conditions. Third and most importantly, an increase in $\lambda_s$ results in a strong reduction of $J^{\rm tot}_s$ below $T_{\rm c}$ for $\text{X} = \text{vacuum}$, whereas it results in an almost temperature-independent $J^{\rm tot}_s$ below $T_{\rm c}$ for $\text{X} = \text{perfect spin sink}$. The difference in this last behavior can be inferred by using approximate expressions $\lambda_s \coth(L/\lambda_s) \approx \lambda_s^2/L$ and $\lambda_s \tanh(L/\lambda_s) \approx L$ for $L/\lambda_s < 1$, both of which appear in the denominator of Eqs.~(\ref{eq:Js01}) and (\ref{eq:Js02}). 

\begin{figure}[t] 
  \begin{center}
    \scalebox{0.65}[0.65]{\includegraphics{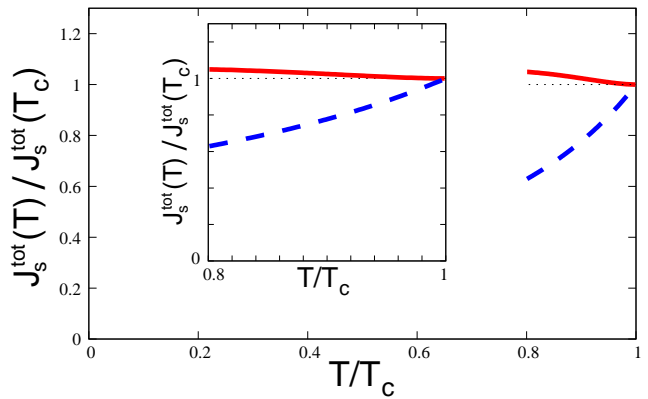}}
  \end{center}
  \caption{Temperature dependence of the spin pumping signal $J_s^{\rm tot}$ in the F/S/X system (see Fig.~\ref{fig_FSsink}) calculated for a temperature window $0.8 < T/T_{\rm c} < 1$. The signal is normalized by its value at $T=T_{\rm c}$. The dashed curve (blue) is the result for the case when X is a vacuum, and the solid curve (red) is the result when X is a perfect spin sink. Inset shows a magnified view. } 
  \label{fig_jtot}
\end{figure}

Figure \ref{fig_jtot} shows the calculated temperature dependence of $J_s^{\rm tot}$ for a region $0.8<T<1$, where the signal is normalized by its value at $T=T_{\rm c}$. The dashed curve (blue) is the result for the case when X is a vacuum, and the solid curve (red) is the result when X is a perfect spin sink. As one can see, the spin pumping signal $J_s^{\rm tot}$ shows a decrease below $T_{\rm c}$ when X is a vacuum, whereas it shows an enhancement below $T_{\rm c}$ when X is a perfect spin sink. This difference in the temperature dependence is explained as follows. When X is a vacuum, the increase in $\lambda_s$ combined with an increase in $D_s$ does not enhance the spin pumping because it just leads to an increase of the spin backflow. By contrast, when X is a perfect spin sink, the increase in $\lambda_s$ combined with an increase in $D_s$ slightly enhance the spin pumping because the spin current arriving at the perfect spin sink is perfectly absorbed. While its dependence on the thickness requires the knowledge of how the superconducting gap $\Delta$ depends on the thickness and thus beyond our scope, in the vicinity of $T_{\rm c}$, the present results are consistent with the experimental finding of Ref.~\cite{Jeon18}.

\section{Discussion and conclusion\label{Sec:VI}}

The main result of the present paper is that the spin relaxation time $\tau_s$ gets longer upon the superconducting transition, as well as the spin diffusion coefficient $D_s$ becomes larger, yielding an increment of the spin diffusion length $\lambda_s$ below $T_{\rm c}$. One of the results that the spin relaxation time grows upon the superconducting transition is consistent with the previous theoretical findings~\cite{Yafet83,Yamashita02}. On the other hand, to the best of our knowledge, the result that the spin diffusion length is increased below $T_{\rm c}$ is different from that of \cite{Yamashita02} and derived for the first time in the present work. In this connection we note that in a previous publication~\cite{Morten05}, the spin diffusion phenomenon in an $s$-wave superconductor was discussed in a spectrally-decomposed way. The present work extends the spectrally-decomposed equation into the integrated equation, which is necessary to discuss hydrodynamic phenomenon such as spin diffusion. We also note that we performed a similar calculation for the case with magnetic impurities, and confirmed that the conclusion of increasing spin diffusion length below $T_{\rm c}$ remains unchanged as long as the condition $\tau_0 \ll \tau_{\rm m}$ is satisfied, where $\tau_{\rm m}$ is the relaxation time by magnetic impurities. 

Given the present result of an increased spin relaxation time below $T_{\rm c}$, it is rather easy to discuss the physics behind the appearance of the coherence peak found in the spin pumping into superconductors~\cite{Inoue17,Yao18,Umeda18}. In short, this is a kind of the inverse of motional narrowing~\cite{Kubo-text,Slichter-text} in that a reduction of the spin-flip scattering rate in superconductors below $T_{\rm c}$ results in a broadening of the magnon damping in the adjacent magnet. Indeed, recalling that the spin conductance characterizing the spin pumping is proportional to the spin relaxation time of the spin sink, i.e., $G_s \propto \tau_s$, in the $\omega \tau_s \ll 1$ limit (see Eq.~(22) in \cite{Maekawa13}), the above analogy to the inverse motional narrowing can be understood. 

To conclude, we have shown microscopically that the spin relaxation time $\tau_s$, the spin diffusion coefficient $D_s$, and the spin diffusion length $\lambda_s$, all are increased just below the superconducting transition temperature $T_{\rm c}$. While our analysis is limited to a temperature region in the vicinity of $T_{\rm c}$, we point out that the present result may provide an alternative explanation for the recent observation of an enhanced spin pumping below $T_{\rm c}$ in a Py/Nb/Pt system~\cite{Jeon18}. 

\acknowledgments 
This work was financially supported by JSPS KAKENHI Grant Number 15K05151. 

\appendix

\section{Coherence peak in the spin pumping signal \label{Sec:App02}}
In this section we calculate the imaginary part of the dynamical susceptibility just below $T_{\rm c}$, in order to see how the coherence peak predicted to appear in the spin pumping signal is derived within the present formalism. We begin with the imaginary part of Eq.~(\ref{eq:chiR01}) and then expand it up to $\Delta^2$: 
\begin{eqnarray}
  \frac{1}{\omega} {\rm Im} \widetilde{\chi}^R_\bmq(\omega) &=&
  \frac{1}{2 \Gamma_{(-)}}
  \frac{{\cal A}^{(0)}}{1-{\cal A}^{(0)}}
  \left( 1+ \frac{\delta {\cal E}}{{\cal A}^{(0)}\left[1-{\cal A}^{(0)}\right]} \right) \nonumber \\
  &\approx&
  \frac{1}{{1}/{\tau_s^{\rm (n)} }+ D_s^{\rm (n)}  q^2}
    \left( 1+ \frac{\delta {\cal E}_0}{1/\tau_s^{\rm (n)} + D_s^{\rm (n)}  q^2} \right), \nonumber \\ 
  \label{eq:ImchiR01}
\end{eqnarray}
where we pick up the contribution dominant in the diffusive limit. The second term is known to be positive from our previous publication~\cite{Inoue17} where the calculation therein is valid to all orders of the superconducting gap $\Delta$. Therefore, the above equation indicates 
\begin{equation}
  \delta {\cal E}_0 =
  \frac{\Delta^2}{\veps_0^2} \frac{1}{1- \gamma} > 0, 
\end{equation}
or equivalently $\gamma < 1$, in order to obtain such a physical result within the present approximate perturbative approach.

\section{Static susceptibility in the superconducting state \label{Sec:App01}} 
In this section, we explain how to derive Eq.~(\ref{eq:Rechi01}). In the static case ($\omega=0$) we have the constraint ${\cal A}{\cal D}+ {\cal B}{\cal C}=0$, such that the summand in Eq.~(\ref{eq:chiQW03}) is written as 
\begin{equation}
  \frac{{\cal A}(1-{\cal D})- {\cal B}{\cal C}}{(1-{\cal A})(1-{\cal D})+{\cal B}{\cal C}}
  = \frac{{\cal A}}{1-{\cal A}-{\cal D}}. 
\end{equation}
Baring the above equation in mind and recalling that in this static case we have ${\cal A}^{(0)}=0$, ${\cal D}^{(0)}= \Gamma_{(-)}/|\widetilde{\veps}_n|$, and $\delta {\cal A}= \Gamma_{(-)} \widetilde{\Delta}^2/|\widetilde{\veps}_n|^3$, the uniform susceptibility is calculated to be 
\begin{eqnarray}
  \chi'_\bmq (\ui \omega_\nu) &=&
  -\frac{\pi N(0)}{\Gamma_{(-)}} T \sum_{\veps_n} 
  \frac{\delta {\cal A}}{1-{\cal D}^{(0)}}, 
\end{eqnarray}
which reproduces Eq.~(\ref{eq:Rechi01}) after adding the normal state contribution.




\end{document}